\documentclass[letterpaper, 10 pt, conference]{ieeeconf}  

\IEEEoverridecommandlockouts  

\usepackage{color, colortbl}
\definecolor{Gray}{gray}{0.9}

\usepackage{mathrsfs}
\usepackage[binary-units]{siunitx}

\usepackage{stfloats}

\usepackage{pstool}

\usepackage{graphicx,psfrag}

\usepackage{cuted}

\usepackage{mathrsfs}
\usepackage{amsthm}

\newtheorem{problem}{Problem}

\usepackage{pdfpages}

\usepackage{hyperref} 

\usepackage{algorithm,algorithmic}

\usepackage{amsfonts}
\usepackage{caption}
\usepackage{graphicx}

\usepackage[utf8]{inputenc}
\usepackage{amssymb}

\usepackage{nomencl} 
\makenomenclature
\usepackage{etoolbox}
\renewcommand{\nomgroup}[1]{%
  \item[\bfseries
  \ifstrequal{#1}{A}{Given Parameters}{%
  \ifstrequal{#1}{B}{Decision variables}{%
  \ifstrequal{#1}{C}{Other Symbols}{}}}%
]}
\usepackage{cite}

\usepackage{amsmath}
\usepackage{algorithmic}

\usepackage{array}

\usepackage{subfigure}

\usepackage{float}

\usepackage{stfloats}

\newcommand{\pushright}[1]{\ifmeasuring@#1\else\omit\hfill$\displaystyle#1$\fi\ignorespaces}
\newcommand{\pushleft}[1]{\ifmeasuring@#1\else\omit$\displaystyle#1$\hfill\fi\ignorespaces}
\makeatother
\newif\ifmargincomments 
\margincommentstrue

\ifmargincomments

\else

\fi

\ifmargincomments

\else

\fi

\begin{document}
%

\title{\LARGE \bf
Incentive-aware Electric Vehicle Routing Problem:\\ a Bi-level Model and a Joint Solution Algorithm
}

%
%

\author{Canqi Yao$^{1}$, Shibo Chen$^{2}$, Mauro Salazar$^{3}$, and Zaiyue Yang$^{2}$

\thanks{ $^{1}$Canqi Yao is with the School of Mechatronics Engineering, Harbin Institute of Technology (HIT), Harbin, 150000, China, and is also with the Shenzhen Key Laboratory of Biomimetic Robotics and Intelligent Systems, Department of Mechanical and Energy Engineering, and the Guangdong Provincial Key Laboratory of Human-Augmentation and Rehabilitation Robotics in Universities, Southern University of Science and Technology (SUSTech), Shenzhen 518055, China.       {\tt\small vulcanyao@gmail.com}}%

\thanks{$^{2}$Shibo Chen and Zaiyue Yang are with the Shenzhen Key Laboratory of Biomimetic Robotics and Intelligent Systems, Department of Mechanical and Energy Engineering, and the Guangdong Provincial Key Laboratory of Human-Augmentation and Rehabilitation Robotics in Universities, Southern University of Science and Technology (SUSTech), Shenzhen 518055, China 
        {\tt\small chensb@sustech.edu.cn, yangzy3@sustech.edu.cn}}%
        
\thanks{$^{3}$ Mauro Salazar is with the Control Systems Technology group, Eindhoven University of Technology (TU/e), Eindhoven, MB 5600, The Netherlands {\tt\small m.r.u.salazar@tue.nl} }        
}

\maketitle

\begin{abstract}
Fixed pickup and delivery times can strongly limit the performance of freight transportation. Against this backdrop, fleet operators can use compensation mechanisms such as monetary incentives to buy delay time from their customers, in order to improve the fleet efficiency and ultimately minimize the costs of operation. To make the most of such an operational model, the fleet activities and the incentives should be jointly optimized accounting for the customers' reactions. Against this backdrop, this paper presents an incentive-aware electric vehicle routing scheme in which the fleet operator actively provides incentives to the customers in exchange of pickup or delivery time flexibility. Specifically, we first devise a bi-level model whereby the fleet operator optimizes the routes and charging schedules of the fleet jointly with an incentive rate to reimburse the delivery delays experienced by the customers. At the same time, the customers choose the admissible delays by minimizing a monetarily-weighted combination of the delays minus the reimbursement offered by the operator. Second, we tackle the complexity resulting from the bi-level and nonlinear problem structure with an equivalent transformation method, reformulating the problem as a single-level optimization problem that can be solved with standard mixed-integer linear programming algorithms. We demonstrate the  effectiveness of our framework via extensive numerical experiments using VRP-REP data from Belgium. Our results show that by jointly optimizing routes and incentives subject to the customers' preferences, the operational costs can be reduced by up to 5\%, whilst customers can save more than 30\% in total delivery fees.
\end{abstract}


%
\IEEEpeerreviewmaketitle

\section{Introduction}

\begin{figure}[t]
\centering
\includegraphics[width=\linewidth]{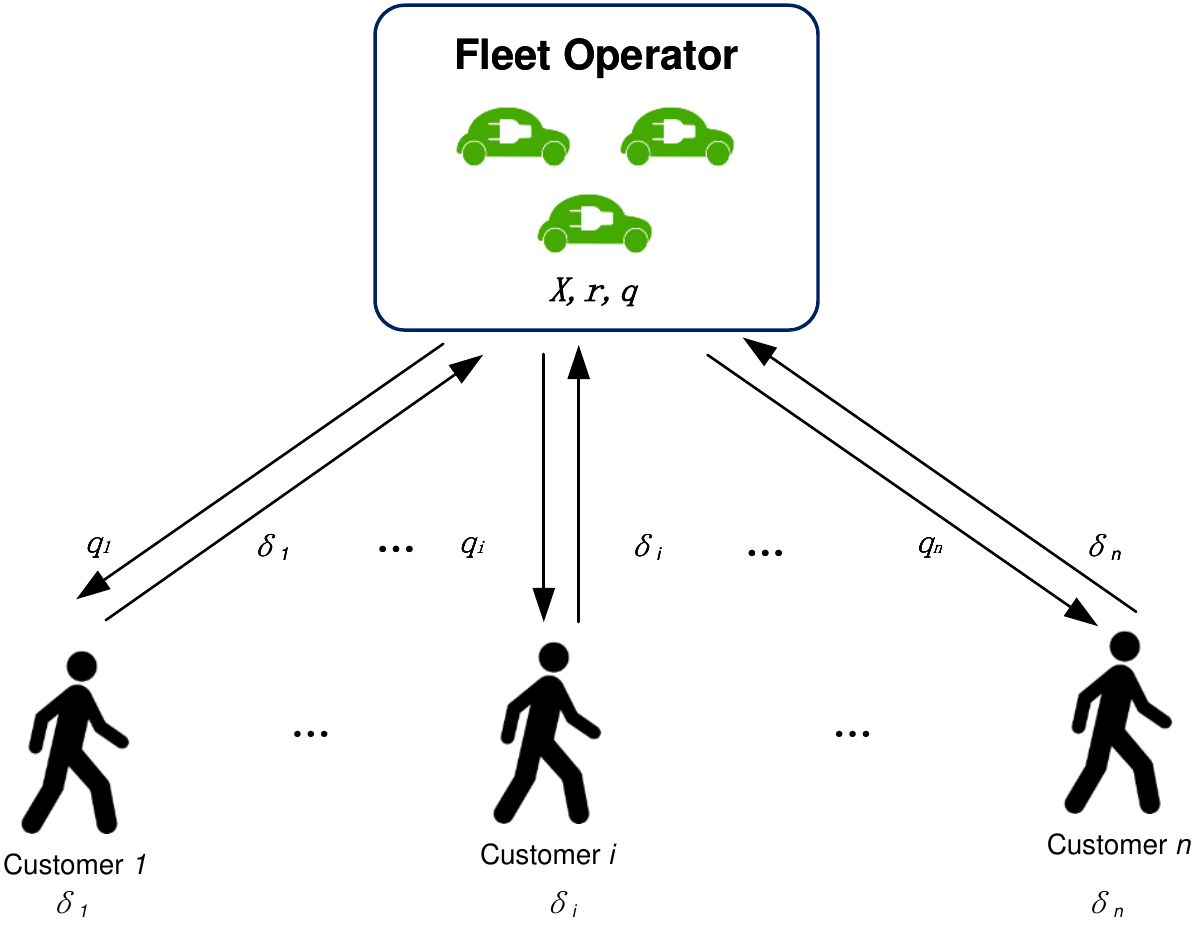}
\caption{Interaction between fleet operator and customers in the bi-level model: The fleet operator optimizes the EV routes $X$ and the charging amounts $r$ jointly with the linear incentive-rates $q$ offered to the customers to compensate for the delays $\delta$; each customer $i$ chooses the admissible delay $\delta_i$ as a function of the incentive rate $q_i$ offered by the operator.}
\label{bilevel}
\end{figure}

In the past decades, online-shopping and e-commerce have been soaring alongside with their customers~\cite{ecommercedata}.
To face such a momentous increase in freight transportation demand, fleets of vehicles have been deployed, including electric vehicles (EVs).
The operators of these fleets (see Fig.~\ref{bilevel}) must strike timely decisions on how to route the vehicles and schedule their charging, whilst adhering to pickup and delivery times.
In fact, strict schedules can limit the operational potential of the fleet, whilst making allowance for some flexibility in terms of transportation delays with respect to the desired schedule can significantly improve the overall operational performance of the fleet.
To this end, fleet operators could buy such delay times from the customers and still improve on the overall revenues.
From the customers' perspective, if enough compensation is offered, it can be beneficial to provide time-flexibility in exchange to a reduction of the transportation fee.
In this context, in order to maximize their revenues, fleet operators should strike the most effective balance between the amount of incentives offered and the benefits stemming from the resulting delays acquired thereto.
Hence, this paper studies the EV routing problem (EVRP) with delays in combination with the incentive design problem accounting for the customers' choices.



\textit{Related literature:} This paper pertains to two research streams: \textit{i)} incentives for mobility and \textit{ii)} the EVRP.
To optimally manage energy demand and promote renewable energy usage in certain areas, researchers have studied incentive schemes to optimize EV charging schedules~\cite{mohsenian2010autonomous,zhou2018incentive,hermans2012incentive,kong2020cloud,sohet2020coupled,narassimhan2018role} and routing~\cite{salazar2021urgency}.
There are few papers~\cite{diaz2021integrated,xiong2020integrated} focused on incentive-based routing and charging, where the fleet operator actively provides incentives to the customers to achieve a better operational performance at the cost of delivery delays.
Aiming to optimize the scheduling scheme, Diaz et al.~\cite{diaz2021integrated} propose an incentive scheme that provides the EV demand-aggregator with more versatility for devising a better operation scheme by choosing the best charging time.
In contrast to fixed-amount incentives,  Xiong et al.~\cite{xiong2020integrated} propose an approach to quantify the personalized monetary incentives and design an algorithm for optimized incentive allocation.
However, the incentive schemes studied in these papers are rather used either for routing or charging applications alone, and are not optimized jointly with the EV routing decisions and charging schedules.

The EVRP has recently received significant attention from the academic community~\cite{basso2021electric}.
To handle the computational complexity of jointly solving the optimal routing and charging problem for delivery EVs, we devised an efficient algorithm based on convex optimization~\cite{yao2021joint}.
Similarly, the vehicle routes and charging operations of a fleet of electric self-driving cars providing on-demand mobility have been optimized with polynomial-time algorithms~\cite{boewing2020vehicle}, and by jointly designing the charging infrastructure via linear programming~\cite{luke2021joint}.
Finally, in our previous work~\cite{yao2021incentive}, we formulated the EVRP with flexible time windows as a bi-level model, which we accurately transformed into a single-level mixed integer nonlinear problem.
However, these papers do not account for the interaction between the fleet operator and the customers.

In conclusion, to the best  of the authors' knowledge, there are no papers jointly solving the electric vehicle routing and charge scheduling problem with the incentive design problem, whilst accounting for the preference of customers on the flexibility in the pickup or delivery time windows.

\textit{Statement of contributions:} Against this backdrop, the contributions of this paper are two-fold: \textit{i)} We present a bi-level model whereby the EV operator solves the EVRP with flexible time-windows and chooses the time-dependent incentives offered to the customers conditional on the delays admitted by the customers; \textit{ii)} we derive a set of exact transformation techniques to transform the original bi-level model to a single-level optimization problem that can be efficiently solved with off-the-shelf mixed-integer linear programming (MILP) algorithms.
Finally, we showcase our framework on a case-study with travel-data from Belgium.


\textit{Organization:} This paper is organized as follows: We elaborate the system models and the mathematical formulation of the bi-level EVRP with time-flexibility in Section~\ref{sec:mathematical}. A set of exact reformulation techniques is proposed to transform the original bi-level model into a single-level MILP in Section~\ref{sec:reformulation}.
Extensive simulation experiments are conducted in Section~\ref{sec:simus} to evaluate the effectiveness of the proposed framework. Finally, Section~\ref{sec:conclusion} concludes the paper with an outlook on future research.

\section{Mathematical Formulation}\label{sec:mathematical}

To characterize the practical scenario, a bi-level optimization model is employed to identify the interaction between customers and EVs operator.
This section presents the customer model and the EVs operator model as optimization problems minimizing the total inconvenience perceived and the fleet operational costs, respectively.
Finally, both models are combined into a bi-level framework.

\subsection{Customer Model}
We assume customers would like to maximize the incentive received, whilst minimizing the inconvenience resulting from the time delay chosen $\delta_j$. 
Similar to~\cite{xu2019model} we quantify the inconvenience of the customers with a piecewise affine function $\mathcal{I}(\delta_j)= \max\limits_{m\in\mathbb{M}=\{1,\cdots,n\} } \gamma_m \delta_j + \chi_m $.
Specifically, given the discount price $q_j$ from the fleet operator, we can define the follower problem for each customer as follows:
\begin{problem}[Customer Model]\label{Model_general_customer}
\begin{equation}\label{P_customer} 
\begin{aligned}
\min_{\delta_j\in\mathbb{R}}\qquad   &  \mathcal{I}(\delta_j)-q_j \delta_j,    \\\text{s.t.}\quad &  0 \leq \delta_j \leq \Bar{\delta}_j
\end{aligned}
\end{equation}    
\end{problem} 

To tackle the difficulties stemming from the non-smooth shape of the objective function, we introduce an auxiliary slack variable $\epsilon_j$, with which we can equivalently reformulate Problem~\ref{P_customer} as a linear problem~\cite{huchette2017nonconvex}:

\begin{problem}[Customer Model Revisited]\label{Model_customer}
    \begin{subequations}
    \begin{align}
  \min_{\delta_j,\epsilon_j\in\mathbb{R}}\quad   &  \epsilon_j-q_j \delta_j ,   \\
\text{s.t.}\quad     &0\leq \delta_j \leq \Bar{\delta}_j \label{customer_bound}\\
& \epsilon_j\geq \gamma_m\delta_j +\chi_m \quad \forall m\in\mathbb{M}.\label{customer_piecewise}
    \end{align}
\end{subequations}
\end{problem}
Thereby, we define the non-negative dual variables $u_j,v_j$ and $\zeta^1_j,\cdots, \zeta^n_j$ associated with the inequality constraints \eqref{customer_bound} and \eqref{customer_piecewise}, respectively.

\subsection{Fleet Operator Model}

We characterize the relation between vehicles and customers with a directed graph as in our previous work~\cite{yao2021joint}.
The transportation network is modeled as a directed graph $G(\mathcal{V},\mathcal{E})$, where $\mathcal{V}=\{v_1,\ldots,v_n\}$ represents start depot, end depot, as well as nodes associated with customers.
The set $\mathcal{E}$ denotes the set of paths with $(i,j)\in\mathcal{E}$ capturing a path from vertex $i$ to $j$. The start depot and end depot are denoted as $v_1$ and $v_n$, respectively.
We define $T_{ij}$ as the travel time between node $i\in \mathcal{V}$ and node $j\in \mathcal{V}$, respectively. The binary variable $x_{ij}^k$  represents whether vehicle $k$ is assigned to traverse path $(i,j)$.

To meet the physical requirements of the charging and transportation system, network flow conservation constraints, battery energy constraints and pickup time constraints are defined as follows: Constraint \eqref{Cons_flow} indicates that all EVs are subject to flow conservation constraint: An EV entering a request node has to exit the same request node, and EVs start at the start depot while returning to the end depot.
  \begin{equation}\label{Cons_flow}
\begin{aligned}
\sum_{j\in\mathcal{V}} & x^k_{i j}-\sum_{j\in\mathcal{V}} x^k_{j i}=b_{i}, \quad \forall  i \in \mathcal{V}; k\in\mathcal{K}\\ &b_{v_1}=1, b_{v_n}=-1, b_{i}=0.
\end{aligned}
\end{equation}
We ensure that each customer can be served at most once as
\begin{equation}\label{Cons_visit}
    \sum_{k\in\mathcal{K}}\sum_{j\in\mathcal{V}} x^k_{ij} \leq 1,
    \quad\forall i\in\mathcal{R}.
\end{equation}
Thereby, some transportation requests may not be served by the fleet if the cost is higher than the benefit. 
Time constraints are given by
\begin{equation}\label{Cons_time}
	\begin{aligned}
		t_{j}   \geq T_{ij}+g_ir^k_{i}+&t_{i} -M(1-x^k_{ij}),  \forall i\in \mathcal{V}\setminus  v_n, \\ &j \in \mathcal{V}\setminus  v_1,k\in\mathcal{K},
	\end{aligned}
\end{equation}
which states that the arrival time at the subsequent request node should be longer than the sum of the arrival times of previous transportation request, charging times, and travel times.
In addition, the time windows are expressed as 
\begin{equation}\label{Cons_time-TW}
	\begin{aligned}
		\tau_{j} \leq	t_{j} \leq  \tau_{j} + \delta_{j} ,  \forall  j \in \mathcal{V}\setminus  v_1.
	\end{aligned}
\end{equation}
The battery energy dynamics of the EVs are characterized by
\begin{equation}\label{Cons_soc}
	\begin{aligned}
		-M (1-x^k_{ij})&\leq -E^k_j+E^k_{i}+r_i^k-e_{i j} x^k_{i j} \leq M (1-x^k_{ij}), \\&\forall i\in\mathcal{V}\setminus  v_n, j\in\mathcal{V}\setminus  v_1,k\in\mathcal{K},
	\end{aligned}
\end{equation}
in which $e_{ij}=\phi d_{ij}$ denotes the amount of energy consumption on edge $ij$, and $\phi$ is the energy consumption per unit distance and $d_{ij}$ is the travel distance from node $i$ to $j$. 
We enforce lower and upper bounds on the battery energy levels of EV $k$ $E_i^k$ as
\begin{equation}\label{Cons_socbound}
    0\leq E^k_i\leq E^k_\mathrm{max}, \quad  i\in \mathcal{V},k\in\mathcal{K},
\end{equation}
where $r_i^k$ is the amount charged at node $i$. 
The initial energy levels of all EVs at the start depot is
\begin{equation}\label{Cons_initialsoc}
E^k_{v_1}=E^k_0 \quad\forall k\in\mathcal{K}.
\end{equation}
Given the time flexibility $\delta_j$ provided by customers, the fleet operator aims at minimizing the operational costs including \textit{(i)} the charging cost $\sum_{k\in \mathcal{K}}\sum_{i\in\mathcal{V}} \sum_{j\in\mathcal{V}}   r^k_{i} p_{i} x^k_{i j}$, \textit{(ii)} the EVs usage cost and delivery revenue
$\sum_{k\in \mathcal{K}}\sum_{i\in\mathcal{V}} \sum_{j\in\mathcal{V}}  c_{i}    x^k_{i j}$, \textit{(iii)} the travel time $\sum_{k\in \mathcal{K}}\sum_{i\in\mathcal{V}} \sum_{j\in\mathcal{V}}    \omega_\mathrm{T} T_{ij}  x^k_{i j}$, \textit{(iv) } the charging time $\sum_{k\in \mathcal{K}}\sum_{i\in\mathcal{V}} \sum_{j\in\mathcal{V}}   \omega_\mathrm{T}  r^k_{i} g_{i}  x^k_{i j}$, and \textit{(v) } the total discount $ \sum_{j\in \mathcal{R}} q_j \delta^*_j \sum_{i\in \mathcal{V}}\sum_{k\in\mathcal{K}} x^k_{ij}$, which represents the cost paid for the customers who provide a flexible delivery time window.
Note that $\omega_\mathrm{T}$ is the value of time. In addition, $c_i$ denotes a unified cost vector which captures both the vehicle usage fee $c_{v}$ and the negative delivery revenue $D_i$\footnote{Since the fleet operator aims to minimize the operation cost, the delivery revenue $D_i$ from the customer is set as a negative value.}:
\begin{equation}\label{Cons:para}
    c_i=\left\{
		\begin{array}{ll}
			{D_i,} & {\text { if } i\in\mathcal{R}} \\ 
			{c_{v},} & {\text { if } i=v_1}
		\end{array}
		\right.
\end{equation}

In conclusion, the fleet operator problem is stated as follows:
\begin{problem}[Operator model]\label{Model_operator}
     \begin{equation}\notag\label{obj}
\begin{aligned}
\min\limits_{x^k_{i j}\in\mathbb{B}, r_i^k, q_j , t_j \in\mathbb{R}}  &\sum_{k\in \mathcal{K}}\sum_{i\in\mathcal{V}} \sum_{j\in\mathcal{V}}   (r^k_{i} p_{i}+c_{i}   + \omega_\mathrm{T} T_{ij}  \\& +  \omega_\mathrm{T}  r^k_{i} g_{i}  ) x^k_{i j} +  \sum_{j\in \mathcal{R}} q_j \delta^*_j \sum_{i\in \mathcal{V}}\sum_{k\in\mathcal{K}} x^k_{ij}\\ 
\centering &\text{s.t. }\eqref{Cons_flow}-\eqref{Cons:para}.
  \end{aligned}
\end{equation}
\end{problem}

The proposed model can be readily applied to capture not only the delivery of goods, but also to the transportation of people.

\subsection{Joint Operator and Customers Optimization Model}
With the customer model and fleet operator model in place, we formulate the problem of jointly optimizing the EV routes and charging schedules with the incentives as the following bi-level problem:
\begin{problem}[Joint optimization model of operator and customers]\label{Model_bilevel}
  \begin{equation}\notag
	\begin{aligned}
		\min\limits_{x^k_{i j}\in\mathbb{B}, r_i^k, q_j , t_j,\delta_j\in\mathbb{R} }& \sum_{k\in \mathcal{K}}\sum_{i\in\mathcal{V}} \sum_{j\in\mathcal{V}}   (r^k_{i} p_{i}+c_{i}   + \omega_\mathrm{T} T_{ij}  \\& +  \omega_\mathrm{T}  r^k_{i} g_{i}  ) x^k_{i j} +  \sum_{j\in \mathcal{R}} q_j \delta^*_j \sum_{i\in \mathcal{V}}\sum_{k\in\mathcal{K}} x^k_{ij}\\
\text{s.t. }&\delta^*_j \in \arg\min_{\delta_j,\epsilon_j\in\mathbb{R}}  \Big\{ \epsilon_j-q_j \delta_j ,   0\leq\delta_j \leq \Bar{\delta}_j, \\&\epsilon_j\geq \gamma_m\delta_j+\chi_m, \forall m\in\mathbb{M} \Big\}, \forall j \in \mathcal{R}\\
&\text{and}\quad \eqref{Cons_flow}-\eqref{Cons:para}.
\end{aligned}
\end{equation}
\end{problem}

\section{Mathematical Reformulation of the Bi-level Optimization Model}\label{sec:reformulation}
Considering that even for the simplest bi-level optimization (e.g., a convex bi-level optimization problem), there is no polynomial time algorithm that can obtain an optimal solution\cite{dempe2015bilevel}, it is unpractical to directly solve Problem~\ref{Model_bilevel}. Instead, we will leverage the KKT conditions of the customers optimization problem to equivalently transform Problem~\ref{Model_bilevel} as a single-level mixed integer nonlinear programming (MINLP) problem. Thereafter, to further reduce the computational complexity of such MINLP, we devise a set of linearization methods to accurately reformulate the resultant MINLP as a MILP.

\subsection{Customer Problem Transformation}
Considering the NP-hardness of this bi-level model, by exploiting the structure information of Problem~\ref{Model_bilevel}, we represent the follower problem as a linear problem via its KKT conditions:
\begin{subequations}\label{Optimality Con}
\begin{align}
&-q_j+u_j-v_j+\sum_{m\in\mathbb{M}}\gamma_m\zeta^m_j=0,\forall j\in\mathcal{R} \label{KKT_stationary_delta}\\
&1-\sum_{m\in\mathbb{M}}\zeta^m_j=0,\forall j\in\mathcal{R} \label{KKT_stationary_epsilon}\\
& 0\leq u_j \perp (\Bar{\delta}_j-\delta_j) \geq 0, \forall j\in\mathcal{R}   \label{KKT_complementary_u}\\
 &0\leq v_j \perp \delta_j \geq 0,\forall j\in\mathcal{R} \label{KKT_complementary_v}\\
&0\leq \zeta^m_j \perp (\epsilon_j- \gamma_m\delta_j -\chi_m ) \geq 0, \forall j\in\mathcal{R}, \forall m\in\mathbb{M}, \label{KKT_complementary_zeta}
\end{align}
\end{subequations}
where the complementarity operator is denoted by $ \perp $. The stationarity conditions are specified in \eqref{KKT_stationary_delta} and (\ref{KKT_stationary_epsilon}), while the primal feasibility, dual feasibility, and complementary conditions are shown in (\ref{KKT_complementary_u}), (\ref{KKT_complementary_v}), and (\ref{KKT_complementary_zeta}). As a result, Problem~\ref{Model_bilevel} can equivalently be reformulated as follows:
\begin{problem}[Single-level MINLP]\label{Model_single_level_NLP}
   \begin{equation}\notag
	\begin{aligned}
		\min\limits_{x^k_{i j}\in\mathbb{B}, r_i^k, q_j , t_j,\delta_j\in\mathbb{R} }& \sum_{k\in \mathcal{K}}\sum_{i\in\mathcal{V}} \sum_{j\in\mathcal{V}}   (r^k_{i} p_{i}+c_{i}   + \omega_\mathrm{T} T_{ij}  \\& +  \omega_\mathrm{T}  r^k_{i} g_{i}  ) x^k_{i j} +  \sum_{j\in \mathcal{R}} q_j \delta^*_j \sum_{i\in \mathcal{V}}\sum_{k\in\mathcal{K}} x^k_{ij}\\
		&
\text{s.t. }\eqref{Cons_flow}-\eqref{Optimality Con}.
	\end{aligned}
\end{equation}
\end{problem}
Considering that there are still bi-linear terms and complementarity constraints in the objective function and constraints of Problem~\ref{Model_single_level_NLP}, respectively, we devise a set of exact linearizations to rewrite such nonlinear terms in a mixed-integer linear fashion.

\subsection{Linearization of the Nonlinear Terms}
To reduce the computation complexity of Problem~\ref{Model_single_level_NLP}, which is an MINLP problem, we devise a set of equivalent linearization approaches, \textit{(i)} linearizing the nonlinear complementary constraint by introducing auxiliary binary variables, and \textit{(ii)} leveraging strong duality\cite{wei2014energy} to exactly linearize the nonlinear term $q_j\delta_j$.

\subsubsection{Linearization of the Complementary Constraints} Due to the nonlinear and nonconvex nature of complementary constraints (\ref{Optimality Con}c), (\ref{Optimality Con}d), and (\ref{Optimality Con}e),  which render the optimization problem extremely hard to solve, we propose to linearize them by introducing auxiliary binary variables $\psi^1_j,\cdots,\psi^{n+2}_j$ and a sufficiently large constant $M$, yielding the following disjunctive constraints~\cite{fortuny1981representation}:

\begin{subequations}\label{Optimality Con linearized}
\begin{align}
&
\left.
\begin{aligned}
&0\leq\Bar{\delta}_j-\delta_j \leq M \psi^{n+1}_j\\
&0\leq u_j \leq M (1- \psi^{n+1}_j)
\end{aligned}
\right\}  \forall j\in\mathcal{R}
\\&
\left.
\begin{aligned}
	&0\leq \delta_j \leq M \psi^{n+2}_j\\
	&0\leq v_j \leq M (1- \psi^{n+2}_j)
\end{aligned}
\right\}  \forall j\in\mathcal{R}\\
& \left.
\begin{aligned}
 &0\leq \epsilon_j-\gamma_m\delta_j-\chi_m \leq M \psi^{m}_j\\
 &0\leq \zeta^m  \leq M (1-\psi^{m}_j)
\end{aligned}
\right\}
 \forall j\in\mathcal{R},m\in\mathbb{M}. 
\end{align}
\end{subequations}

\subsubsection{Linearization of the Objective Function}
Note that there are still nonlinear terms in the objective function of Problem~\ref{Model_single_level_NLP}. Inspired by the work of \cite{wei2014energy}, we see that there is no duality gap between the lower-level customer problem and its dual problem, as the former is convex. Therefore, the objective value of the customer problem  and  its  dual  problem are equal at the optimum due to strong duality, implying
\begin{equation*}
\begin{aligned}
&\epsilon_j-q_j \delta_j = -u_j\Bar{\delta}_j + \sum_{m\in\mathbb{M}}\zeta_m \chi_m \\ \Longrightarrow\quad& q_j \delta_j= \epsilon_j + u_j\Bar{\delta}_j-\sum_{m\in\mathbb{M}}\zeta_m \chi_m. 
\end{aligned}
 \end{equation*}




In addition, to handle the bilinear terms $\omega_\mathrm{T} r_i^k g_i x^k_{ij}$, $r_i^k p_i x^k_{ij}$  and $ (\epsilon_j + u_j\Bar{\delta}_j-\sum_{m\in\mathbb{M}}\zeta_m \chi_m)x^k_{ij} $ in the objective function of Problem~\ref{Model_single_level_NLP}, the continuous auxiliary variables $\eta^1_{ijk},\eta^2_{ijk},\eta^3_{j}$ are introduced as
\begin{subequations}\label{obj_linearized}
	\begin{align}
		&\eta^1_{ijk}\geq\omega_\mathrm{T}r^k_ig_i-M(1-x^k_{ij}),\forall i\in\mathcal{V},j\in\mathcal{V},k\in\mathcal{K}\\
		&\eta^2_{ijk}\geq r^k_ip_i-M(1-x^k_{ij}),\forall i\in\mathcal{V},j\in\mathcal{V},k\in\mathcal{K} \\
		& \begin{aligned}
     	\eta^3_{j}\geq \epsilon_j + u_j\Bar{\delta}_j-&\sum_{m\in\mathbb{M}}\zeta_m \chi_m-M(1-\sum_{k\in\mathcal{K}}\sum_{i\in\mathcal{J}} x^k_{ij})\\&\forall j\in\mathcal{V}.
		\end{aligned}
	\end{align}
\end{subequations}

\subsubsection{Equivalent Reformulation of Problem~\ref{Model_single_level_NLP}}
For convenience, we group all optimization variables in  $X=\{\{x^k_{i j},\eta^1_{ijk},\eta^2_{ijk}\}_{i,j\in\mathcal{V}}^{k\in\mathcal{K}}, \{q_j,\delta_j,\epsilon_j, u_j,v_j,\zeta^1_j,\cdots,\zeta^n_j, \psi^1_j,\cdots,\\ \psi^{n+2}_j,\eta^3_{j}\}_{j\in\mathcal{V}} , \{r_i^k\}_{i\in\mathcal{V}}^{k\in\mathcal{K}} \}.$ With the aforementioned exact linearization approaches, Problem~\ref{Model_single_level_NLP} can be equivalently reformulated as the following single-level mixed-integer linear optimization problem, which can be efficiently solved by off-the-shelf MILP solvers~\cite{bliek1u2014solving}:
  \begin{problem}[Single-level MILP]\label{Model_single_level_MILP}
      \begin{equation}\notag
\begin{aligned}
&\min\limits_{X} \sum_{k\in \mathcal{K}}\sum_{i\in\mathcal{V}} \sum_{j\in\mathcal{V}} ( \eta^1_{ijk}+ \eta^2_{ijk}+ \eta^3_{j}  + ( \omega_\mathrm{T} T_{ij} + c_{i}) x^k_{i j} )
  \end{aligned}
\end{equation}
$$\textit{s.t.}\qquad \eqref{Cons_flow}-\eqref{Cons:para},\eqref{KKT_stationary_delta},\eqref{KKT_stationary_epsilon}, \eqref{Optimality Con linearized},\eqref{obj_linearized}. $$
  \end{problem}

\section{Numerical Results}\label{sec:simus}
This section showcases our optimization framework on an example of package delivery. We present the numerical data used in Section~\ref{sec:pars} below.
\subsection{Parameter Settings}\label{sec:pars}
We use VRP-REP data for Belgium\footnote{\url{http://www.vrp-rep.org/datasets/item/2017-0001.html}}, which records the geographic coordinates of 1000 nodes, out of which we randomly select a finite amount of delivery nodes.
The battery capacity and charging power are selected as \SI{90}{\kWh}, \SI{150}{\kW}(fast charging), and \SI{22}{\kW}(slow charging)\footnote{\url{https://www.tesla.com/}}. 
In addition, the energy consumption rate and average speed of EVs are set as \SI[per-mode=symbol]{0.24}{\kWh\per\km} and \SI[per-mode=symbol]{60}{\km\per\hour}, respectively. According to the electricity price of Belgium\footnote{\url{https://www.globalpetrolprices.com/Belgium/electricity_prices/}}, the charging price at a fast charging station is \SI[per-mode=symbol]{0.299}{\$\per kWh}, whilst it is \SI[per-mode=symbol]{0.129}{\$\per kWh}\footnote{\url{https://www.currentautomotive.com/a-guide-to-teslas-supercharging-network/}} at a slow charging station.
The usage cost of an EV $c_{v}$ is set as \SI[per-mode=symbol]{199}{\$}.
To characterize a practical scenario, we sample the revenue gained when serving a request from a normal distribution with a mean of \SI[per-mode=symbol]{9.05}{\$} and a standard deviation of \SI{5}{\$}\footnote{\url{https://www.fedex.com/en-us/shipping/one-rate.html}}.
The inconvenience function is modeled as a two-segment piecewise affine function with $\gamma_{1,2} = \{ 0,1.5\}\si{\$\per\hour} $ and  $\chi_{1,2} = \{0.01,-0.01\}\si{\$} $ for every customer. Note that this framework can be easily extended to the heterogeneous case (i.e., different customers have different sensitivities).

We run the numerical experiments 50 times, and show the average results.
All optimization problems are implemented with Python 3.7 on a computer with an Intel Core i9-10980XE processor with 36 CPUs of \SI{3.00}{\giga\hertz} and \SI{64}{\giga\byte} of memory.
To obtain an optimal routing and charging scheme of EVs, we solve Problem~\ref{Model_single_level_MILP} with the Gurobi Python interface gurobipy~\cite{pedroso2011optimization}.
The resulting computation times are between \SI{0.02}{\second} and \SI{7200}{\second} with an average value of \SI{340}{\second} and a standard deviation of \SI{1238}{\second}.



  


\subsection{Numerical Experiments}
To demonstrate the advantage of using incentives for time-flexibility, we compare the solution of Problem~\ref{Model_single_level_MILP} with different values of the maximum delay $\bar{\delta}_j$ (which we set to be the same for each customer) and sensitivity $\gamma_2$ with the solution of the incentive-free problem detailed in Appendix~\ref{app:noincentives}.

\subsubsection{Impact of Different Time-flexibility Levels}\label{Si:time Flexibility on operation cost} 
\begin{figure}[t]
\centering
 \includegraphics[width=\linewidth]{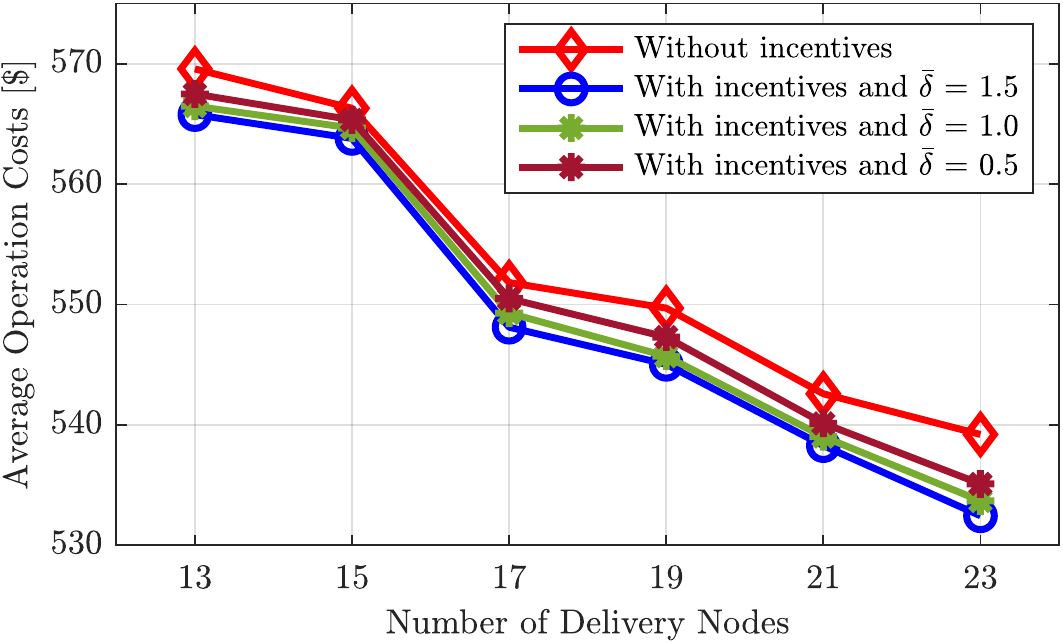}
\caption{Average operation cost over 50 experiments with different time-flexibility levels. }
\label{OperationCostReduction_delta}
\end{figure}

\begin{figure}[t]
\centering
\includegraphics[width=\linewidth]{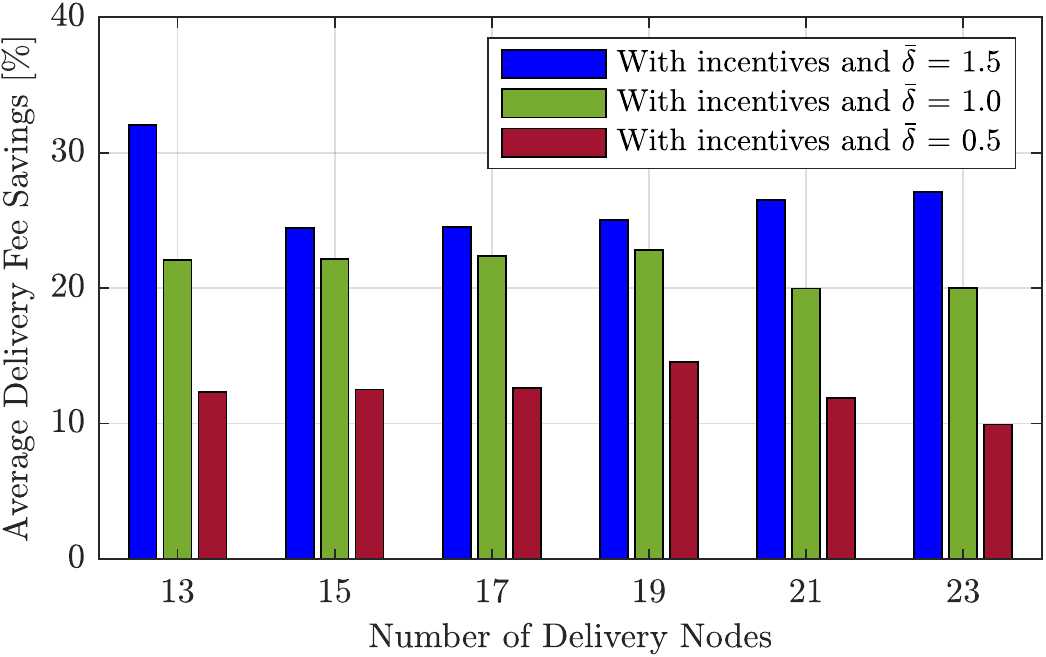}
\caption{Average delivery fee over 50 experiments with different time-flexibility levels. }
\label{DeliveryFeeReduction_delta}
\end{figure}

In order to investigate the impact of different length of time flexibility on the reduction of operation cost, we conduct 3 sets of experiments by increasing the value of $\bar{\delta}$ from \SI{0.5}{\hour} to \SI{1.5}{\hour}.
As shown in Fig.~\ref{OperationCostReduction_delta}, the operation cost becomes smaller with increasing values of $\bar{\delta}$, revealing that the operation cost can be further reduced when customers allow for more time flexibility.
In turn, a larger value of time flexibility will enable customers to obtain a larger incentive and ultimately a larger delivery fee saving, as depicted in Fig.~\ref{DeliveryFeeReduction_delta}. Thereby, the maximum operation cost reduction is up to $5\%$ (this case happens in the specific simulation scenario, and is not shown in Fig.~\ref{OperationCostReduction_delta}), whilst the largest average delivery fee saving exceeds $30\%$.


\subsubsection{Impact of Different Time-sensitivity Levels}\label{Si:time sensitivity on operation cost}
\begin{figure}[t]
\centering
\includegraphics[width=\linewidth]{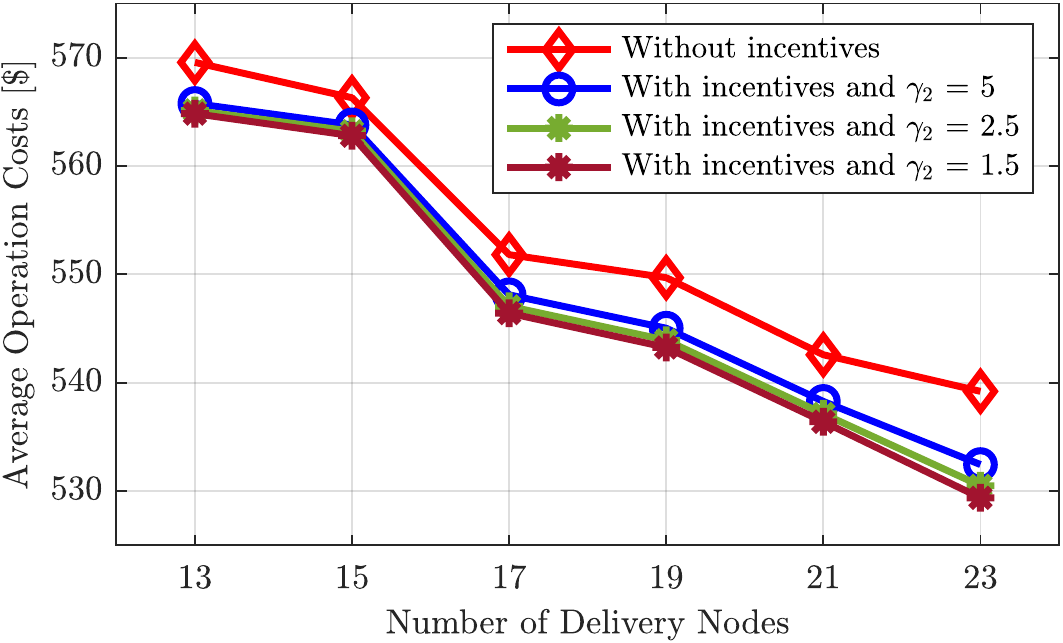}
\caption{Average operation costs over 50 experiments with different values of $\gamma_2$. }
\label{OperationCostReduction_gamma}
\end{figure}

\begin{figure}[t]
\centering
\includegraphics[width=\linewidth]{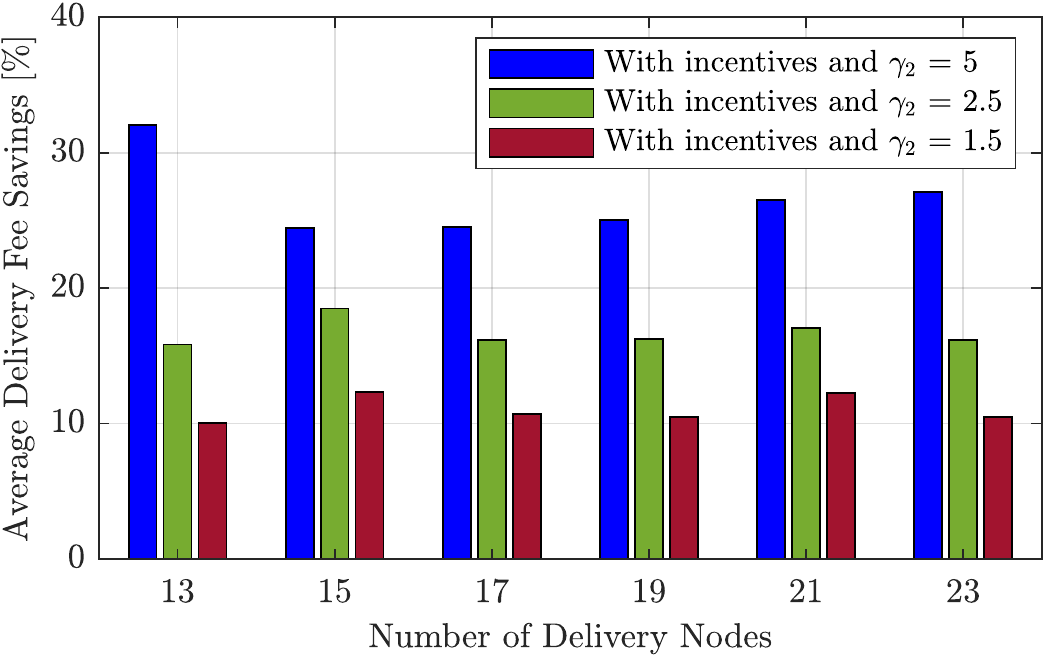}
\caption{Average delivery fee savings over 50 experiments with different values of $\gamma_2$. }
\label{DeliveryFeeReduction_gamma}
\end{figure}

To evaluate the impact of the different degree of time sensitivity of customers on the operation cost reductions, we carry out three sets of experiments with $\gamma_2 \in \{1.5,2.5,5\}\si{\$\per\hour}$.
Fig.~\ref{OperationCostReduction_gamma} shows that the value of the average operation cost becomes larger with increasing values of $\gamma_2$.
As delays become more inconvenient for the customers, the fleet operator needs to pay them a higher incentive. However, even in the extreme case, it is always beneficial for the operator to pay such incentives and leverage the flexibility of variable time windows.
The same reasoning applies to the delivery fee savings.
As shown in Fig.~\ref{DeliveryFeeReduction_gamma}, the higher the value of $\gamma_2$, the higher the average fee savings, which can exceed $30\%$.

\section{Conclusion}\label{sec:conclusion}
This paper devised a bi-level optimization model to frame and solve the incentive-aware electric vehicle routing problem with flexible time windows.
To handle the complexity stemming from the bi-level and nonlinear problem structure, we reformulated the problem as a mixed integer linear problem that could be solved with off-the-shelf algorithms.
Our numerical results showed that allowing for flexible time windows and jointly optimizing the incentives offered to the customers in exchange of delay time could improve the fleet operator overall costs by up to $5\%$, whilst customers could save over $30\%$ of the total delivery fees.
In particular, a larger time-sensitivity and/or maximum delay allowed by the customers always resulted in larger customers' savings. Thereby, even in the most extreme scenarios, it was always beneficial for the fleet operator to pay incentives and leverage the flexibility of the resulting delivery windows.

This work can be extended as follows:
First, we are interested in including the time-of-use electricity price in our model. Second, we would like to further improve the computational efficiency of our algorithm and implement it in real-time.
Finally, we would like to study more complex customers' models under uncertainty.


\section*{Acknowledgments}
We thank Dr.\ I.\ New for proofreading this paper.\\
\indent This work was supported in part by the National Key Research and Development Program of China under Grant 2019YFB1705401; in part by the Natural Science Foundation of China under Grant 61903179 and 61873118; in part by the Science, Technology and Innovation Commission of Shenzhen Municipality under Grant RCBS20200714114918137, 20200925174707002 and ZDSYS20200811143601004.

\appendices

\section{Electric Vehicle Routing Problem without Incentives}\label{app:noincentives}

 \begin{equation}\label{EVRP}\notag
\begin{aligned}
\min\limits_{x^k_{i j}, r_i^k, q_j , t_j }  \sum_{k\in \mathcal{K}}\sum_{i\in\mathcal{V}} \sum_{j\in\mathcal{V}}   &(r^k_{i} p_{i}+c_{i}   + \omega_\mathrm{T} T_{ij}   +  \omega_\mathrm{T}  r^k_{i} g_{i}  ) x^k_{i j}    
  \end{aligned}
\end{equation}
\begin{subequations}
    \begin{align}
        \text{s.t.~}  &\sum_{j\in\mathcal{V}}  x^k_{i j}-\sum_{j\in\mathcal{V}} x^k_{j i}=b_{i}, \quad \forall  i \in \mathcal{V}; k\in\mathcal{K}
        \\ &\quad\qquad b_{v_1}=1, b_{v_n}=-1, b_{i}=0,\notag
        \\&\sum_{k\in\mathcal{K}}\sum_{j\in\mathcal{V}} x^k_{ij} \leq 1,\quad\forall i\in\mathcal{R} 
    \\& t_{j}   \geq T_{ij}+g_ir^k_{i}+t_{i} -M(1-x^k_{ij}),  \forall i\in \mathcal{V}\setminus  v_n, 
    \\ &\qquad\quad j \in \mathcal{V}\setminus  v_1,k\in\mathcal{K}
	\notag 
	\\&\tau^\mathrm{L}_{j} \leq	t_{j} \leq  \tau^\mathrm{U}_{j} ,  \forall  j \in \mathcal{V}\setminus  v_1
	\\&-M (1-x^k_{ij})\leq -E^k_j+E^k_{i}+r_i^k-e_{i j} x^k_{i j}  \\&\qquad\leq M (1-x^k_{ij}),\forall i\in\mathcal{V}\setminus  v_n, j\in\mathcal{V}\setminus  v_1,k\in\mathcal{K}\notag
	\\& 0\leq E^k_i\leq E_{k,\mathrm{max}}, \quad  i\in \mathcal{V},k\in\mathcal{K}    
	\\& E^k_{v_1}=E^k_0, \quad\forall k\in\mathcal{K}.
    \end{align}
\end{subequations}

\bibliographystyle{IEEEtran}  
\bibliography{references}









\end{document}